\documentclass[twocolumn,preprintnumbers,aps,prc]{revtex4-1}

\usepackage[T1]{fontenc}
\usepackage{graphics}
\usepackage{amssymb}
\usepackage{amsfonts}
\usepackage{amsmath}
\usepackage{fancyhdr}
\usepackage{here}
\usepackage{hyperref}
\usepackage{epsfig}
\usepackage{epstopdf}
\usepackage{color}

\begin{document}
	
	\title{Covariant interacting Hadron-Resonance Gas model}
	\author{T.~Steinert}%
	\email{Thorsten.Steinert@theo.physik.uni-giessen.de}
	\affiliation{%
		Institut f{\"u}r Theoretische Physik, %
		Universit\"at Giessen, %
		35392 Giessen, %
		Germany %
	}
	
	\author{W.~Cassing}
	\affiliation{%
		Institut f{\"u}r Theoretische Physik, %
		Universit\"at Giessen, %
		35392 Giessen, %
		Germany %
	}
	\date{\today}
	\begin{abstract}
		The Hadron-Resonance Gas (HRG) approach  - used to model hadronic matter at small baryon  potentials $\mu_B $  
and finite temperature $T$ - is extended to finite and large chemical potentials by introducing interactions between baryons in 
line with  relativistic mean-field theory defining an interacting HRG (IHRG). Using lattice data for $\mu_B=0$ as well as 
information on the nuclear equation of state at $T=0$ we constrain the attractive and repulsive interactions of the IHRG  such 
that it reproduces the lattice equation of state at $\mu_B=0$ and the nuclear equation of state at $T=0$ and finite $\mu_B$. The 
formulated covariant approach is thermodynamically consistent and allows us to provide further information on the phase boundary 
between hadronic and partonic phases of strongly interacting matter by assuming constant thermodynamic potentials.
  \\
		
		\par
		% explanations PACS [OL]:
		PACS: 12.40.Ee, 21.30.Fe, 21.65.Mn, 24.10.Pa
		%
		%12.40.-y 	Other models for strong interactions
        %12.40.Ee 	Other models for strong interactions, Statistical models
		%21.30.Fe 	Nuclear forces, Forces in hadronic systems and effective interactions
		%21.65.-f 	Nuclear matter 
		%21.65.Mn 	Equations of state of nuclear matter
		%24.10.Pa   Nuclear reactions: general, Thermal and statistical models
		%25.75.-q 	Relativistic heavy-ion collisions
		%25.75.Ag 	Global features in relativistic heavy ion collisions
		%25.75.Dw 	Particle and resonance production
		
		%
		%QGP 12.38.Mh
		%Chirality, particle physics, 11.30.Rd
		%Relativistic heavy-ion collisions, 25.75.-q
		%
	\end{abstract}
	\maketitle
	\section{Introduction}
The phase diagram of matter is one of the most important subjects in physics since it also has important implications on chemistry
and biology. Accordingly, the phase diagram of strongly interacting matter is a topic of utmost interest since decades and
substantial experimental and theoretical efforts have been invested to shed light on this issue. The early universe went through
different phases at practically vanishing baryon chemical potential $\mu_B$ when expanding to its present size. Relativistic and
ultra-relativistic heavy-ion collisions nowadays offer the unique possibility to study some of these phases, in particular a
quark-gluon plasma (QGP) phase and its phase boundary to the hadronic one. Lattice Quantum-Chromo-Dynamics (lQCD) calculations 
suggest that at vanishing baryon chemical potential $(\mu_B=0)$ there is a crossover phase transition from hadronic to partonic 
degrees of freedom 
\cite{Aoki:2006br,Borsanyi:2010bp,Borsanyi:2012uq,Borsanyi:2015waa,Petreczky:2012gi,Petreczky:2012fsa,Ding:2015ona} for the 
deconfinement phase transition as well as for the restoration of chiral symmetry. However, at some finite baryon chemical 
potential the crossover might turn  to  a first-order phase transition implying a critical endpoint in the QCD phase diagram 
\cite{Senger:2011zza}. Since lattice calculations so far suffer from the fermion-sign problem, no first-principles information on 
the phase boundary can be extracted from lQCD at large $\mu_B$, whereas at low $\mu_B$ Taylor expansions of the thermodynamic 
potential (in powers of $\mu_B/T$) provide an alternative solution as demonstrated in Refs. \cite{Bazavov:2017dus,Karsch:2013fga}. 
Accordingly, heavy-ion reactions at Relativistic Heavy-Ion Collider (RHIC) and Large Hadron Collider (LHC) energies show no 
evidence for a phase transition and experimental studies at much lower bombarding energies are needed to explore the high  $\mu_B$ 
region of the phase diagram. To this end new facilities - such as the Facility for Antiproton and Ion Research (FAIR) and the 
Nuclotron-based Ion Collider fAcility (NICA) - have been planned and are presently under construction.

The QCD phase diagram has two regions that are relatively well known, while the rest is more or less unknown. These two regions
are the temperature axis at $\mu_B=0$, which can be studied by lQCD calculations, and the region of $T=0$, which is
described by nuclear physics. Usually one constructs models that can only be applied to one of the two cases, but we intend to set
up an approach that is able to describe both regions simultaneously and to allow for more stringent extrapolations on unknown 
areas of the phase diagram. A common model used in the first regime is the Hadron-Resonance Gas (HRG) model that treats  hadrons 
as a gas of non-interacting particles. This model works at vanishing chemical potential but fails for the description of nuclear 
matter due to the lack of repulsive and attractive interactions. The latter  are included in relativistic mean-field theories 
(RMFTs) whose interactions are based on meson-exchange potentials. While these models can describe infinite nuclear matter with 
the right properties of the binding energy, they fail for the QCD equation of state at vanishing chemical potential $\mu_B$. We 
will combine both approaches in the following to set up a model that is consistent with lQCD $(\mu_B \approx 0,\ T>0)$ and the 
nuclear equation of state $(T \approx 0,\ \mu_B>0)$ while using only hadronic degrees of freedom. Of course, nothing can be 
predicted for the partonic phase at high temperature or chemical potential, but we aim at defining a phase boundary between the 
hadronic and partonic phases by approaching it from the hadronic side.

It is commonly expected that the HRG gives a good description of hadronic 'QCD matter' at moderate chemical potentials due to the 
success of the statistical hadronization model in describing particle ratios from relativistic heavy-ion collisions. This model 
assumes that the medium - created in a heavy-ion collision - equilibrates (to some extent). The system will then continue to 
interact until it becomes too dilute and freezes out. The particle yields then are fixed by the temperature and the chemical 
potential at the chemical freeze out (apart from the volume, which drops out in particle ratios). This simple model can describe 
the particle abundances for various collision energies 
\cite{Cleymans:1990nz,Cleymans:1992zc,BraunMunzinger:2003zd,Andronic:2005yp} from Alternating Gradient Synchrotron (AGS) up to LHC
energies. It becomes even more precise if one assumes additional corrections for non-equilibrium effects 
\cite{Letessier:2005qe,Petran:2013lja,Becattini:2003wp,Becattini:2005xt}. The model is applicable also for $p+p$
\cite{Vovchenko:2015idt} and even $e^+e^-$ collisions \cite{Becattini:2010sk} and was also applied to the production of 
hypernuclei in Ref. \cite{Andronic:2010qu}. Furthermore, the HRG equation of state and susceptibilities were compared to state of 
the art lQCD calculations in Refs. 
\cite{Borsanyi:2011sw,Bazavov:2012jq,Borsanyi:2013bia,Bazavov:2014pvz,Borsanyi:2012cr,Borsanyi:2010cj,Bellwied:2015lba,
Bellwied:2013cta} and it was found that the ideal HRG leads to a satisfying description of the thermodynamics for temperatures
below $T \approx 170 \ \text{MeV}$. The quality of the description is improved if one includes an exponential increasing mass 
spectrum \cite{NoronhaHostler:2012ug}, as predicted by Hagedorn, and repulsive interactions 
\cite{Vovchenko:2014pka,Begun:2012rf,Andronic:2012ut,Vovchenko:2016rkn,Samanta:2017yhh,Huovinen:2017ogf}.

On the other hand, the main application for relativistic mean-field theories is the calculation of ground-state properties of
infinite nuclear matter and finite nuclei. The model works well for spherical and deformed nuclei 
\cite{Gambhir:1989mp,Ring:1996qi}, and  can also be used to investigate neutron star properties 
\cite{Klahn:2006ir,Sugano:2016jcb}, in particular the mass-radius relation. Furthermore, the RMFT provides the basis for 
non-equilibrium transport approaches when applied to heavy-ion collisions at lower energies or merely the hadronic phase 
\cite{Cassing:1999es,Linnyk:2015rco}.

This work is organized as follows: In Sec. \ref{sec:Hadron-Resonance Gas} we recapitulate the equation of state from the HRG at
vanishing chemical potential while in Sec. \ref{sec:Nuclear equation of state} we recall the framework of the relativistic
mean-field model and discuss the nuclear equation of state at vanishing temperature. In Sec. \ref{sec:IHRG} we will combine both
approaches in a thermodynamic consistent manner and define the covariant interacting hadron resonance gas (IHRG) approach, which
will allow us to define a phase boundary in the $(T, \mu_B)$ plane from the hadronic side. This study is summarized in Sec.
\ref{sec:summary}.

\section{Reminder of the Hadron-Resonance Gas (HRG) Model}
\label{sec:Hadron-Resonance Gas}

The most frequently used model for the thermodynamics of hadrons at finite temperature $T$ and baryon chemical potential $\mu_B$
is the HRG. The approach is based on the work of Dashen, Ma and Bernstein \cite{Dashen:1969ep}, who found that one can describe 
the thermodynamics of a system of particles, which interact through resonant scatterings, by including the resonances as stable
particles in the partition sum. This is always possible if the spectral widths of the resonances are small compared to the
temperature $\gamma \ll T$. As an example we quote the interacting pion gas, which is thermodynamically equivalent to a free gas of
pions and $\rho$-mesons \cite{Welke:1990za,Venugopalan:1992hy}. The HRG generalizes this approach to all possible hadrons such 
that the thermodynamic potential for the hadronic system $\Omega_{HRG}$ is given by the sum over all stable hadrons and all known 
hadronic resonances,
\begin{equation}
 \label{eq:HRG}
 \Omega_{HRG}(T,\mu)=\sum_{\text{Had}} \Omega_0(T,\mu,m_i)+\sum_{\text{Res}} \Omega_0(T,\mu,m_i),
\end{equation}
without mutual interactions,
\begin{equation}
	\label{I1}
	\Omega_0(T,\mu,m_i)=-\frac{d_i}{6 \pi^2} \int_0^{\infty} dp \ \frac{p^4}{\sqrt{p^2+m^2_i}} \ n_{B/F}.
\end{equation}
In Eq. (\ref{I1}) $n_{B/F}$ denotes the Bose/Fermi distribution while $d_i$ is the particle degeneracy.
The approach incorporates attractive interactions (for the dynamical formation of resonances) but discards repulsive
interactions that describe the short-range repulsion between the hadrons. However, the effects from short-range repulsion can be
introduced by assuming a finite volume of the particles \cite{Rischke:1991ke,Yen:1997rv,Kapusta:1982qd,Hagedorn:1980kb} which is
excluded in the thermodynamic analysis and thus leads to an increase of the pressure $P$ (for fixed particle number, cf. Eq. (\ref{VDW})). The model 
presented in Refs. \cite{Rischke:1991ke,Yen:1997rv} assumes the same volume for each particle such that the excluded volume is 
proportional to the total particle density. The approach in Refs. \cite{Kapusta:1982qd,Hagedorn:1980kb} assumes the excluded 
volume for each particle to be proportional to its energy, which leads to a limiting energy density similar to the Hagedorn model 
(with a maximum temperature \cite{Hagedorn:1965st}). A non-relativistic version of this approach is the van der Waals model with 
the pressure $P$ and internal energy $U$ given by,
\begin{equation}
\label{VDW}
 P=\frac{N T}{V-b N}-a\frac{N^2}{V^2}, \qquad U=\frac{3}{2} N T-a \frac{N^2}{V},
\end{equation}
where $b$ characterizes the excluded volume and $a$ the strength of the attractive interaction. The repulsive interactions, which
are incorporated in excluded volume models, are quite different from repulsive interactions  originating from  vector
mesons exchange as, for example, in the Walecka model \cite{Walecka:1974qa}. In the latter case the strength of the vector
repulsion is proportional to the net particle density and thus vanishes for $\mu_B=0$. On the other hand, the excluded volume
repulsion is proportional to the total particle density (or pressure) and is finite also at vanishing chemical potential.

The only parameters in (\ref{eq:HRG}) are the masses of the hadrons, which are usually taken as the vacuum masses. In principle,
the model is parameter free, but one has to decide on the amount of 'non-interacting particles' to include. The most fundamental
hadrons are the spin-$1/2$ baryons and $0^-$ mesons as well as the spin-$3/2$ baryons and the $1^-$ mesons as resonances. All of
these hadrons are important for the dynamics of heavy-ion collisions and have to be included to achieve a reasonable description
of finite temperature QCD in the confined hadronic phase. In the context of chiral symmetry restoration at high temperature and/or
baryon density other hadrons also play an important role, i.e. the chiral partners of opposite parity, e.g. the  $a_1$-meson and
the $N(1440)$ and $N(1535)$ baryon or the  scalar $0^+$ mesons. Among the $0^+$ mesons the most prominent is the $f_0(500)$ or 
$\sigma$-meson with a mass of $400 - 550 \ \text{MeV}$ \cite{Pelaez:2015qba}, which is now established as a particle and contained 
in the latest version  of the Particle Data Group (PDG) \cite{Olive:2016xmw}. The other mesons in the scalar $0^+$-multiplet are 
the  $f_0(980)$ and the $a_0(980)$, which are both listed by the PDG, as well as the strange $\kappa(720)$ meson. However, there 
is no consensus whether to include the $\sigma$ and the $\kappa$ in HRG models or not since calculations for the thermodynamic 
potential of an interacting pion gas in terms of experimental phase shifts show that the attractive pressure contribution from the 
scalar $\sigma$-mesons gets exactly canceled by the repulsive isotensor channel \cite{Venugopalan:1992hy,Broniowski:2015oha}; the 
same happens also for the $\kappa$-mesons. However, in these calculations the vacuum phase shifts have been employed. On the 
other hand,  effective hadron field theories suggest that the $\sigma$-meson is much stronger affected by finite temperature or 
chemical potential effects than the pions and $\rho$-mesons 
\cite{Tripolt:2013jra,Bernard:1987im,Hatsuda:1986gu,Wergieluk:2012gd}; accordingly vacuum  phase shifts should no longer
be valid for temperatures above $T \approx 100 \ \text{MeV}$. Arguments in favor of the scalar mesons come from the statistical
hadronization model in Ref. \cite{Andronic:2008gu}, where the $\sigma$-meson was explicitly included to improve the description of
the $K^+/\pi^+$ ratio that was observed experimentally in central Au+Au (Pb+Pb) collisions. Furthermore,  it was shown in Ref.
\cite{Pelaez:2015qba} that a neglect of the $0^+$-mesons in the thermodynamic partition sum is inconsistent with respect to
causality and unitarity.

\begin{figure}[t]
	\centering
	\includegraphics[width=0.90\linewidth]{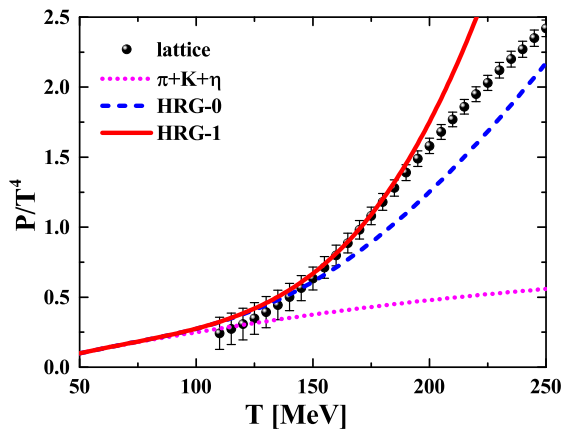}
	\caption{The dimensionless pressure $P/T^4$ for a HRG with varying amounts of hadrons as a function of the temperature
		$T$. The dotted magenta line considers only pions, kaons and  $\eta$'s, the dashed blue line is the basic HRG-0 and the full red
		line shows the pressure for all hadrons listed by the particle data group \cite{Agashe:2014kda} with a mass below $2.6 \
		\text{GeV}$ (HRG-1). The lQCD results are taken from the
		Wuppertal-Budapest collaboration from Ref. \cite{Borsanyi:2013bia}.}
	\label{pic:HRG_EoS_vs_T}
\end{figure}

\begin{figure}[t]
	\centering
	\includegraphics[width=0.90\linewidth]{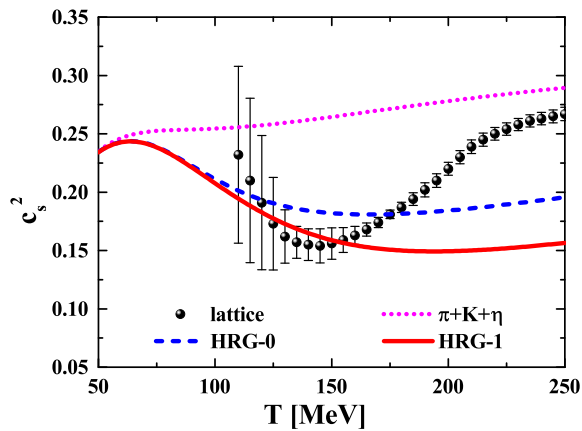}
	\caption{The speed of sound squared $c_s^2$ for a non-interacting hadron gas as a function of temperature. The meaning of
		the lines is the same is in Fig. \ref{pic:HRG_EoS_vs_T}. The lQCD results are taken from the 
Wuppertal-Budapest collaboration from Ref. \cite{Borsanyi:2013bia}.}
	\label{pic:HRG_cs2_vs_T}
\end{figure}

Although the $0^-$, $0^+$ and $1^-$ mesons, the spin-$1/2$ and spin-$3/2$ baryons together with e.g. the $a_1$, $N(1440)$ and
$N(1535)$ are the most basic hadronic degrees of freedom required to describe the hadronic medium, they do not produce the 
required pressure at higher temperatures (above $\sim$ 150 MeV)  to describe the lQCD equation of state. It is therefore necessary 
to include additional hadrons and a standard choice is to incorporate all hadrons listed by the Particle Data Group
\cite{Olive:2016xmw,Agashe:2014kda} with a mass below a certain threshold $M_{max}$. For illustration we show in Fig.
\ref{pic:HRG_EoS_vs_T} the dimensionless pressures $P/T^4=-\Omega_{HRG}/T^4$ at vanishing chemical potential calculated with different numbers of particles  in comparison  to the recent lQCD data from the Wuppertal-Budapest collaboration 
\cite{Borsanyi:2013bia}. We denote the HRG with the most basic hadrons (quoted above) as HRG-0 and show also the result when including  all 
hadrons from the PDG \cite{Agashe:2014kda} with a mass below $2.6 \ \text{GeV}$ (HRG-1). A compact list of the hadrons, but 
without the $\sigma$ and the $\kappa$-meson can be found in Ref. \cite{Albright:2014gva}. We see that at low temperatures the 
system is dominated by the lightest mesons, i.e. pions, kaons and the $\eta$-meson, which describe the equation of state up to $T 
\approx 120 \ \text{MeV}$. In general the equation of state is meson dominated and baryons contribute only above $T \approx 140 \ 
\text{MeV}$. The basic HRG-0 describes the lQCD data only up to $T \approx 150 \ \text{MeV}$ and underestimates the pressure at 
higher temperatures. The additional hadrons in HRG-1 provide the necessary pressure to describe the equation of state up to $T 
\approx 180 \ \text{MeV}$ within the error bars of the lQCD result. However,  the speed of sound defined by
\begin{equation}
 c_s(T)=\sqrt{\frac{\partial P(T)}{\partial E(T)}},
\end{equation}
with the energy density
\begin{equation}
E(T)=\frac{1}{2 \pi^2} \sum_i d_i \int_0^{\infty} dp \ p^2 \sqrt{p^2+m^2_i} \ n_{B/F}
\end{equation}
turns out to fail the lQCD results. This is demonstrated in Fig. \ref{pic:HRG_cs2_vs_T} where the speed of sound (squared) 
$c_s^2$ - for the different HRG versions - is compared with the lQCD data taken from the same simulation as the pressure
\cite{Borsanyi:2013bia}. Whereas the lQCD data show a clear minimum for $T \approx 150 \ \text{MeV}$,  the HRG versions do not
show a substantial increase for higher temperature regardless of the hadron  content. Any inclusion of further (heavier)
resonances leads to an overall decreasing speed of sound for $T > 150 \ \text{MeV}$.

Extending the HRG to even heavier resonances with masses beyond $2.6 \ \text{GeV}$ has almost no effect on the pressure because 
additional hadronic degrees of freedom have only small effects on the equation of state. However, a comparison between the HRG 
and lQCD data shows that the experimentally established hadrons are not sufficient to describe strangeness fluctuations 
\cite{Bazavov:2014xya}, which indicates that a full description of QCD (above about temperatures of 150 MeV) in terms
of hadronic degrees of freedom requires an even stronger interaction.

\section{Relativistic mean-field theory}
\label{sec:Nuclear equation of state}

As mentioned above, an increase of the pressure can be achieved by an excluded volume in the van der Waals model that mimics the
effects of short-range repulsive forces. The latter are naturally included in covariant effective theories with baryons and 
mesons by the massive vector mesons. We recall that the limit of $T=0$ and finite chemical potential $\mu_B$ is known as the 
"infinite nuclear matter" limit, a scenario found in the interior of all larger atomic nuclei, where the nuclear density is 
almost constant. The binding energy of nuclear matter $E_B/A=E/\rho_B-m_N$ has a minimum 
of $E_B/A \approx -16 \ \text{MeV}$ at normal nuclear density $\rho_0 \approx $ 0.16 fm$^{-3}$ to reproduce the
stable nuclear matter inside finite atomic nuclei. For $T=0$ the Fermi-distribution function becomes
\begin{equation}
\label{Fermi}
 n_F=\frac{1}{e^{\frac{\omega_p-\mu_B}{T}}+1}=\Theta(\mu_B-\sqrt{p^2+m^2})=\Theta(p_F-p),
\end{equation}
while the distribution function for anti-fermions and the Bose-distribution functions for mesons vanish. In Eq. (\ref{Fermi}) the
momentum $p_F$ is the Fermi-momentum which specifies the largest occupied momentum state; at $T=0$ it fixes the baryon density
via
\begin{equation}
\label{ROB}
 \rho_B=4 \int \frac{d^3p}{(2 \pi)^3} n_F(\omega_p)=\frac{4}{6 \pi^2} p_F^3
\end{equation}
and it is therefore common to describe nuclear matter properties in terms of the density $\rho_B$. In (\ref{ROB}) a degeneracy
factor of 4 has been introduced (for protons and neutrons with two spin projections). Note that without anti-baryons the net-baryon
density and the total baryon density are the same; this changes for $T \neq 0$.

Since only baryons with a mass larger than the chemical potential $\mu_B$ can populate the system at $T=0$, the conventional
HRG can not describe the nuclear equation of state and in particular the minimum at $\rho_0$. Baryons other than nucleons can
appear only at very large (energy) densities that most likely have to be described by partonic degrees of freedom. The HRG -
without interactions -  reduces  essentially to a gas of nucleons and thus fails not only at high $T$ and $\mu_B = 0$, but also
for $T=0$.

The interactions between nucleons conventionally are assumed to be mediated by meson exchange and are described by relativistic
mean-field theories \cite{Walecka:1974qa,Serot:1984ey,Vretenar:2005zz,Serot:1997xg}. One usually includes isoscalar interactions -
mediated by the scalar $\sigma$-meson and the vector $\omega$-meson - and isospin-dependent interactions that are mediated by the
$\rho$-meson and the $\delta$-meson. The $\sigma$-meson describes the attractive part of the nucleon-nucleon interaction while the
$\omega$-meson is responsible for the short-range repulsion. The $\rho$ and the $\delta$-meson are important for asymmetric
nuclear matter and neutron star physics, but give no contribution in isospin symmetric matter. Since this is approximately the
case for the hot and dense medium created in heavy-ion collisions, we can neglect them in the following. The Lagrangian of
relativistic mean-field theory then reads,
\begin{align}
\label{eq:RMF_lag}
 \mathcal{L}&=\mathcal{L}_B+\mathcal{L}_M+\mathcal{L}_{int},\\
 \mathcal{L}_B&=\bar{\Psi} \left(i \gamma_{\mu} \partial^{\mu}-M \right) \Psi,\\
 \mathcal{L}_M&=\frac{1}{2} \partial_{\mu} \sigma \partial^{\mu} \sigma-U(\sigma)-\frac{1}{4} F_{\mu \nu}
F^{\mu \nu}+O(\omega^{\mu} \omega_{\mu}),\\
\label{eq:QHD-lag-int}
 \mathcal{L}_{int}&=\Gamma_{\sigma N}(\bar{\Psi},\Psi) \bar{\Psi} \sigma \Psi-\Gamma_{\omega N}(\bar{\Psi},\Psi) \bar{\Psi}
\gamma^{\mu}
\omega_{\mu} \Psi,
\end{align}
where the functions $U(\sigma)$ and $O(\omega^{\mu} \omega_{\mu})$ describe selfinteractions of the $\sigma$- and the
$\omega$-fields that are introduced to incorporate non-linear density dependences as arising from Dirac-Brueckner calculations.
The couplings $\Gamma_{\sigma N}$ and $\Gamma_{\omega N}$ are not constants but can be functions of the nucleon field
\cite{Fuchs:1995as,Typel:1999yq,Hofmann:2000vz} in order to better comply with results from Dirac-Brueckner calculations, too. To
preserve Lorentz invariance the couplings  have to be Lorentz scalars. The easiest way to ensure this is to write them as a 
function of a density $\Gamma(\bar{\Psi},\Psi)=\Gamma(\hat{\rho}_0)$, which is a Lorentz scalar itself. Two physical reasonable 
choices are $\hat{\rho}_0=\bar{\Psi} \Psi$ and $\hat{\rho}_0=\bar{\Psi} u_{\mu} \gamma^{\mu} \Psi$, where $u_{\mu}$ is the 
four-velocity with $u_{\mu} u^{\mu}=1$. The first one is denoted by scalar density dependence (SDD) and will lead to a dependence 
on the scalar density $\rho_s$, the second one is called vector density dependence (VDD) and will lead to a dependence on the 
baryon density $\rho_B$. It has been shown, that the application of the VDD gives better results when applied to finite nuclei in 
Refs. \cite{Fuchs:1995as,Typel:1999yq}. Furthermore, the density dependence of the couplings has the advantage that one can 
parametrize a realistic nucleon-nucleon interaction - as obtained from Dirac-Brueckner (DB) calculations - with less numerical 
effort \cite{Typel:1999yq,Hofmann:2000vz}, which allows us to apply DB calculations also to finite systems. We point out that only 
with density-dependent couplings we can reproduce the nuclear equation of state and the lQCD equation of state simultaneously in 
the same approach. Note, that for the thermodynamics one can describe the non-linear density dependence of the DB-interactions 
either with density-dependent couplings or with non-trivial mesonic selfinteractions.

Since the Lagrangian \eqref{eq:RMF_lag} is too complicated to be solved on the many-body level,  we will use the mean-field
approximation, where the meson fields are no longer independent degrees of freedom but determined by their expectation values.
When evaluating the equations of motion one ends up with the following two coupled equations \cite{Fuchs:1995as}, which
have to be solved simultaneously:
\begin{align}
 \label{eq:selfcon1}
 \frac{\partial U}{\partial \sigma}&=\Gamma_{\sigma N}(\rho_0) \  \rho_s(T,\mu^*,m^*),\\
 \label{eq:selfcon2}
 \frac{\partial O}{\partial \omega}&=\Gamma_{\omega N}(\rho_0) \ \rho_B(T,\mu^*,m^*).
\end{align}
with
\begin{align}
 \rho_s= d \int \frac{d^3p}{(2 \pi)^3} \frac{m^*}{\omega^*_p}
\left( n_F(T,\mu^*,\omega^*_p)+n_F(T,-\mu^*,\omega^*_p) \right),\\
 \rho_B=d \int \frac{d^3p}{(2 \pi)^3}\left(  n_F(T,\mu^*,\omega^*_p)- n_F(T,-\mu^*,\omega^*_p) \right).
\end{align}
Here $d$ is the degeneracy of the fermion field, which in case of nucleons is $d=4$. The density in the couplings is now the
normal ordered expectation value of the density $\rho_0=\langle : \hat{\rho}_0 : \rangle$. The distribution functions $n_F$ depend
on $\omega^*_p=\sqrt{\textbf{p}^2+m^{*2}}$ with the effective mass
\begin{align}
 m^* &=m_N-\Sigma^s=m_N-\Sigma^{s(0)}-\Sigma^{s(r)}\\
 & =m_N-\Gamma_{\sigma N}(\rho_0) \sigma-\Sigma^{s(r)} \notag
\end{align}
and on the effective chemical potential
\begin{equation}
\label{eq:mueff}
 \mu^*=\mu-\Sigma^0=\mu-\Sigma^{0(0)}-\Sigma^{0(r)}=\mu-\Gamma_{\omega N}(\rho_0) \omega-\Sigma^{0(r)},
\end{equation}
which both get affected by the interactions with the mesons. The mass $m^*$ gets modified by the scalar selfenergy $\Sigma^s$
that originates from the interactions with the $\sigma$-meson and the chemical potential gets modified by the vector selfenergy
$\Sigma^0$ that originates from the interactions with the $\omega$-meson. The selfenergies are split into a conventional
$\Sigma^{(0)}$ and a rearrangement selfinteraction $\Sigma^{(r)}$, which arises from the density dependence of the couplings.
Their actual forms depend on the choice of $\rho_0$. If the couplings are independent from the fields and just constants, the
rearrangement selfenergies vanish, but otherwise they are mandatory for thermodynamic consistency (cf. Ref. \cite{Fuchs:1995as}).
In this work we will employ constant scalar couplings and VDD vector couplings. The rearrangement selfenergies for this case read
\begin{equation}
\label{eq:rearrangement}
 \Sigma^{s(r)}=0, \qquad \qquad \Sigma^{0(r)}=\frac{\partial \Gamma_{\omega N}}{\partial \rho_B} \omega \rho_B.
\end{equation}
The pressure and the energy density of the density-dependent relativistic mean-field model are given by
\begin{align}
 P&= P_0(T,\mu^*,m^*)+\Sigma^{0(r)} \rho_B -\Sigma^{s(r)}\rho_s-U(\sigma)+O(\omega),\\
\label{eq:RMF_E}
E&= E_0(T,\mu^*,m^*) +\Sigma^{s(r)}\rho_s + \Sigma^{0(0)}\rho_B + U(\sigma) - O(\omega),
\end{align}
where $P_0$ and $E_0$ are the pressure and energy density for a non-interaction particle evaluated for the effective quantities
$\mu^*$ and $m^*$. The model is thermodynamic consistent as long as the selfconsistent equations of motion \eqref{eq:selfcon1} 
and \eqref{eq:selfcon2} are fulfilled.

\begin{figure}[t]
	\centering
	\includegraphics[width=0.90\linewidth]{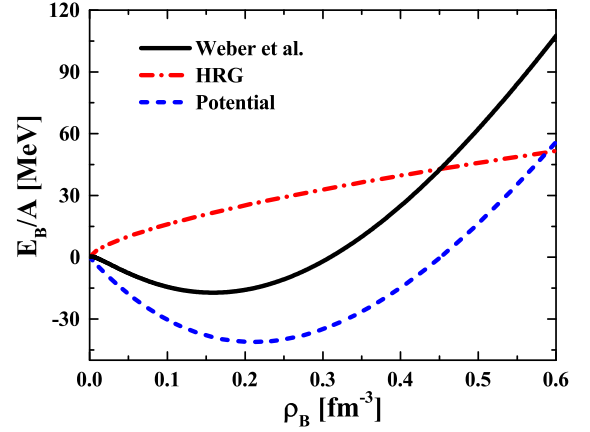}
	\caption{The binding energy per particle $E_B/A$ as a function of the nuclear density $\rho_B$. The solid black line
		shows the equation of state from Ref. \cite{Weber:1992qc} (as a benchmark) while the dash-dotted red line is the 
binding energy per nucleon for non-interacting nucleons. The dashed blue line shows the potential energy contribution 
that is missed by the non-interacting nuclear model. }
	\label{pic:EBT0}
\end{figure}

The binding energy per nucleon (for $T=0$) as a function of $\rho_B$ has been the subject of extensive studies for decades and
is not so well known for densities above about 3 $\rho_0$. Note, however, that for a baryon density of 3 $\rho_0$ one obtains an
energy density $\sim$ 0.5 GeV/fm$^3$, which corresponds to the critical energy density in case of $\mu_B=0$. In this work we will
use the binding energy per nucleon from Ref. \cite{Weber:1992qc}, which is consistent with microscopic Dirac-Brueckner
calculations and the experimentally known momentum dependence of the nucleon-nucleus optical potential (full black line in Fig.
\ref{pic:EBT0}).
We recall that the covariant approach in Ref. \cite{Weber:1992qc} is energy-momentum conserving and most importantly
also thermodynamically consistent.  The conventional HRG result is shown by the red dash-dotted line in Fig. \ref{pic:EBT0} and
shows no binding as pointed out before.  A suitable parametrization of relativistic mean-field theory, which reproduces the
binding energy for densities up to 0.6 fm$^{-3}$, is given by:
\begin{align}
 \Gamma_{\sigma N}&=9.26, & \Gamma_{\omega N}&=10.59, \notag\\
 m_{\sigma}&=550 \ \text{MeV}, & m_{\omega}&=782 \ \text{MeV},\\
 B&=5.1 \ \text{fm}^{-1},  & C&=9.8, \notag
\end{align}
with the meson selfenergies given by
\begin{align}
\label{eq:U_sigma}
 U(\sigma)&=\frac{1}{2} m_{\sigma}^2 \sigma^2+\frac{1}{3} B \sigma^3+\frac{1}{4}C \sigma^4,\\
\label{eq:O_omega}
 O(\omega)&=\frac{1}{2} m_{\omega}^2 \omega^2.
\end{align}
The couplings  here are taken as independent from the nucleon fields and the rearrangement selfenergies thus vanish. The minimum 
of this binding energy is $E_B/A=-16.1 \ \text{MeV}$ at a density of $\rho_0=0.164 \ \text{fm}^{-3}$.

\begin{figure}[t]
	\centering
	\includegraphics[width=0.90\linewidth]{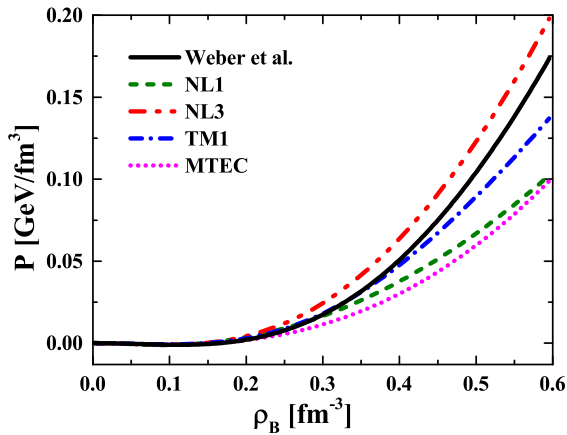}
	\caption{The pressure for symmetric nuclear matter as a function of the nuclear density $\rho_B$ from Ref. 
\cite{Weber:1992qc} (black solid line) and for different parametrizations of the relativistic mean-field model from Refs. 
\cite{Lang:1991qa,Sugahara:1993wz,Tatsumi:2005qn}.}
	\label{pic:nuc_P_RMF}
\end{figure}

We show in Fig. \ref{pic:nuc_P_RMF} the pressure from the equation of state from Ref. \cite{Weber:1992qc} (black solid line) together with other 
parametrizations for relativistic mean-field models taken from Refs. \cite{Lang:1991qa} (NL1 and NL3), \cite{Sugahara:1993wz} 
(TM1) and \cite{Tatsumi:2005qn} (MTEC). The sets TM1 and MTEC are both stiff enough to allow for neutron stars with two solar 
masses (when extended by the isovector $\rho$ exchange) as shown in Ref. \cite{Sugano:2016jcb}.

%\color{black}

We close the discussion of the nuclear equation of state with a short remark on the thermodynamic potential. So far we have shown
the equation of state at $T=0$ as a function of the density $\rho_B$ and not of the chemical potential $\mu_B$ since it is more 
convenient to describe the system in terms of the Fermi momentum $p_F$ ($\rho_B  \sim p_F^3$). It is therefore natural to express 
all thermodynamic quantities as a function of $\rho_B$, which implies, that we are not working in a grand-canonical ensemble but 
in a canonical ensemble. In this case the thermodynamic potential is the free energy, $F=U-TS$, which is equal to the internal 
energy $U$ for $T=0$. The pressure of the system in this special case follows from the relation,
\begin{equation}
\label{eq:EB_to_P}
 P=\rho_B^2 \frac{\partial}{\partial \rho_B} \left( \frac{E}{\rho_B} \right)=\rho_B^2 \frac{\partial}{\partial \rho_B}
\left( E_B/A \right),
\end{equation}
and is no longer proportional to the thermodynamic potential (as in case of the grand-canonical ensemble). As one can see from
Eq. \eqref{eq:EB_to_P} this implies also negative pressures if the binding energy per nucleon decreases with $\rho_B$, i.e. for
all densities below the saturation density $\rho_0$ (cf. Fig. \ref{pic:nuc_P_RMF}). The system for $\rho_B<\rho_0$ is unstable,
since the grand-canonical thermodynamic potential is larger than the vacuum. Another consequence of the canonical nature of the 
nuclear equation of state is the uncertainty in the chemical potential, which is no longer a natural variable. The system is 
described by the effective chemical potential $\mu^*$ \eqref{eq:mueff}, i.e. the real chemical potential gets shifted by the 
repulsive interaction. It might therefore be impossible to define the nuclear equation of state as a function of the chemical 
potential in a unique manner.

\section{Interacting Hadron-Resonance Gas (IHRG)}
\label{sec:IHRG}

In this Sec. we introduce interactions in the covariant hadronic model by constraining the Lagrange density at vanishing chemical
potential ($\mu_B=0$)  by the most recent equation of state from the Wuppertal-Budapest collaboration \cite{Borsanyi:2013bia} and
at vanishing temperature ($T=0$)  by the nuclear equation of state from Ref. \cite{Weber:1992qc}. At finite temperature mesons
will appear and interact with other mesons and baryons through resonant scatterings, which we describe in terms of the HRG by
including several important resonances as non-interacting particles. We restrict the particles here to the most basic hadrons
summarized by the list HRG-0 in Sec. \ref{sec:Hadron-Resonance Gas}. We, furthermore, incorporate meson-exchange interactions in
terms of relativistic mean-field models, which introduces additional attractive interactions as mediated by the $\sigma$-meson,
which will account for the missing higher resonances in the HRG. The $\sigma$ and the $\omega$-meson appear in this model also as
non-interacting particles, which may seem as a double counting, but the non-interacting contribution plays the role of an
$s$-channel resonant-scattering amplitude. This channel - which is neglected
in the mean-field limit - is missing in the meson-exchange model and the "exchange particles" appear only in the $t$ channel. The nucleons appear  in both channels,
but in their case it is important to omit the non-interacting contribution in the HRG to avoid a true double counting. The
thermodynamic potential of the interacting Hadron-Resonance Gas (IHRG) then is defined by the sum of the relativistic mean-field
model and the regular HRG without nucleons,
\begin{equation}
 \Omega_{IHRG}=\Omega_{RMF} + \Omega_{HRG} - \Omega_{0,N}.
\end{equation}
All other thermodynamical quantities, which follow from $\Omega_{IHRG}$ by differentiation, then are also just the sum of the HRG
and RMF contributions, but without the non-interacting nucleons.

For the description of the nuclear equation of state it is not necessary to extend the model towards more interacting particles,
since additional baryons will only appear for very large densities $\rho_B \geq 2 - 3 \ \rho_0$
\cite{Boguta:1982rsg,Li:1997yh,Drago:2013fsa}. This changes at finite temperature where other baryons populate the system and
start to interact via meson exchange. However, the interaction has to be tuned in such a way, that it does not change the nuclear
equation of state. The inclusion of additional interacting baryons such as hyperons and $\Delta$'s is a frequently discussed question
in the context of neutron-star physics, see Ref. \cite{Drago:2013fsa} and references therein. We will use the findings from this
field to extend the IHRG to include further interacting baryons while the mesons are kept non-interacting. Especially important in
this context - and also for the description of heavy-ion collisions - are the $\Delta$-resonances, which we describe by the
Lagrangian \cite{Boguta:1982rsg,Li:1997yh,Kosov:1997yx},
\begin{align}
\label{eq:RMF_Delta_lag}
 \mathcal{L}_{\Delta}=&\bar{\Psi}_{\Delta \nu}\left(i \gamma_{\mu} \partial^{\mu}-M_{\Delta} \right)
\Psi_{\Delta}^{\nu}\\
&+\Gamma_{\sigma \Delta} ( \hat{\rho}_0 ) \bar{\Psi}_{\Delta \nu} \sigma
\Psi_{\Delta}^{\nu} -\Gamma_{\omega \Delta}(\hat{\rho}_0) \bar{\Psi}_{\Delta \nu} \gamma^{\mu}\omega_{\mu} \Psi_{\Delta}^{\nu},
\notag
\end{align}
which is added to the Lagrangian of the relativistic mean-field theory \eqref{eq:RMF_lag}. The couplings $\Gamma_{\Delta}$ may
depend on an arbitrary Lorentz scalar or stay constant. The spinor $\Psi_{\Delta}^{\nu}$ is not a Dirac spinor but a
Rarita-Schwinger spinor with $4 \times 4$ components that describes a spin-3/2 particle \cite{Rarita:1941mf}, however, the
mean-field limit of the theory behaves just like Dirac spinors \cite{Boguta:1982rsg,Li:1997yh}. The selfconsistent equations
\eqref{eq:selfcon1} and \eqref{eq:selfcon2} become
\begin{align}
 \label{eq:RMF_Delta_selfcon1}
 \frac{\partial U}{\partial \sigma}&=\Gamma_{\sigma N}\rho_s^N(T,\mu^*_N,m^*_N)+\Gamma_{\sigma \Delta}
\rho_s^{\Delta}(T,\mu^*_{\Delta},m^*_{\Delta}),\\
 \label{eq:RMF_Delta_selfcon2}
 \frac{\partial O}{\partial \omega}&=\Gamma_{\omega N}\rho_B^N(T,\mu^*_N,m^*_N)+\Gamma_{\omega \Delta}
\rho_B^{\Delta}(T,\mu^*_{\Delta},m^*_{\Delta}).
\end{align}
Here $\rho_s^{\Delta}$ and $\rho_B^{\Delta}$ are the scalar and the particle density for non-interacting $\Delta$-baryons. They
depend on the effective mass $m^*_{\Delta}$ and the effective chemical potential $\mu^*_{\Delta}$ that are defined by
the selfenergies of the $\Delta$'s,
\begin{align}
 \label{eq:RMF_Delta_effective_m_mu}
 m^*_{\Delta}=m_{\Delta}-\Sigma^s_{\Delta}&=m_{\Delta}-\Gamma_{\sigma \Delta} \sigma-\Sigma^{s(r)}_{\Delta},\\
 \mu^*_{\Delta}=\mu-\Sigma^0_{\Delta}&=\mu-\Gamma_{\omega \Delta} \omega -\Sigma^{0(r)}_{\Delta}.
\end{align}
The actual form of the rearrangement selfenergies follows from the density-dependence of the couplings $\Gamma_{\sigma \Delta}$ and $\Gamma_{\omega \Delta}$. The pressure and the energy density of the system - without the HRG contribution - are given by
\begin{align}
 P=& P_0(T,\mu^*_N,m^*_N)+ P_0(T,\mu^*_{\Delta},m^*_{\Delta})+\Sigma^{0(r)}_N \rho_B^N  \\
  &+ \Sigma^{0(r)}_{\Delta} \rho_B^{\Delta} -\Sigma^{s(r)}_N\rho_s^N -\Sigma^{s(r)}_{\Delta}\rho_s^{\Delta} -U(\sigma)+O(\omega) \notag
\end{align}
and
\begin{align}
 E=& E_0(T,\mu^*_N,m^*_N) +E_0(T,\mu^*_{\Delta},m^*_{\Delta}) +\Sigma^{s(r)}_N\rho_s^N \\ 
 & + \Sigma^{s(r)}_{\Delta} \rho_s^{\Delta}+\Sigma^{0(0)}_N\rho_B^N +\Sigma^{0(0)}_{\Delta} \rho_B^{\Delta}+ U(\sigma) - O(\omega) \notag.
\end{align}
The entropy and the particle density are simply given by the non-interacting expressions, but with the respective effective
quantities $\mu^*_{\Delta}$, $m^*_{\Delta}$,
\begin{align}
\label{eq:RMF_delta_S}
 s&=s^N_0(T,\mu^*_N,m^*_N) +s^{\Delta}_0(T,\mu^*_{\Delta},m^*_{\Delta}),\\
\label{eq:RMF_delta_N}
 \rho_B&=n^N_0(T,\mu^*_N,m^*_N) +n^{\Delta}_0(T,\mu^*_{\Delta},m^*_{\Delta}).
\end{align}
The approach is thermodynamically consistent if the selfconsistent equations \eqref{eq:RMF_Delta_selfcon1} and
\eqref{eq:RMF_Delta_selfcon2} are fulfilled. The thermodynamic potential of the IHRG with interacting nucleons and $\Delta$'s is
\begin{equation}
 \Omega_{IHRG}=\Omega_{RMF} + \Omega_{HRG} - \Omega_{0,N}- \Omega_{0,\Delta},
\end{equation}
where $\Omega_{RMF}$ is now the relativistic mean-field theory with nucleons and $\Delta$-baryons. This extension introduces two additional couplings $\Gamma_{\sigma \Delta}$ and $\Gamma_{\omega \Delta}$. As for nucleons
these couplings are not fixed by theory, but one can impose several constraints. The introduction of additional particles such as
$\Delta$'s or hyperons can create a second minimum in the binding energy \cite{Boguta:1982rsg,Li:1997yh}, but since there are no
$\Delta$'s in the ground state of nuclear matter, this minimum can only describe a metastable state. Furthermore, any contribution
from the $\Delta$'s has to vanish at saturation density. There is also some guidance from finite density sum-rules, which show
that the scalar selfenergy of the $\Delta$'s is larger and the vector selfenergy smaller than the corresponding values for the
nucleon selfenergies \cite{Jin:1994vw}. In Ref. \cite{Kosov:1997yx} all these conditions are used to constrain the model in case
of constant couplings. These findings are summarized by:
\begin{equation}
\label{eq:coupling-relations}
 \frac{\Gamma_{\sigma \Delta}}{\Gamma_{\sigma N}} \leq 1.01 \cdot \frac{\Gamma_{\omega \Delta}}{\Gamma_{\omega N}} + 0.38, \qquad
 \frac{\Gamma_{\sigma \Delta}}{\Gamma_{\sigma N}} \geq 1, \qquad  \frac{\Gamma_{\omega \Delta}}{\Gamma_{\omega N}}
\leq 1.
\end{equation}
A simple choice for the couplings - in line with the relations \eqref{eq:coupling-relations} - are the conditions $\Gamma_{\sigma
\Delta}/\Gamma_{\sigma N}=m_{\Delta}/m_N$ and $\Gamma_{\omega \Delta}=\Gamma_{\omega N}$. They are based on the argument that the
$\omega$-meson has a real quark-antiquark structure and the $\sigma$-meson does not \cite{Li:1997yh}. This choice leads to a fixed
ratio of the effective masses $m^*_{\Delta}/m^*_N=m_{\Delta}/m_N$ and equal chemical potentials for both baryons,
$\mu^*_{\Delta}=\mu^*_N$. We will employ this choice whenever we treat the $\Delta$'s as interacting particles.

The generalization to even more interacting baryons is straight forward. Spin-$1/2$ and spin-$3/2$ particles behave equally
in the mean-field limit. We fix the scalar couplings by the ratio of the bare masses and keep the vector couplings identical,
\begin{equation}
\label{eq:coupling_gen_RMF}
 \frac{\Gamma_{\sigma X}}{\Gamma_{\sigma N}}=\frac{m_X}{m_N}, \qquad \qquad \Gamma_{\omega X}=\Gamma_{\omega N}.
\end{equation}
The selfconsistent equations \eqref{eq:RMF_Delta_selfcon1} and \eqref{eq:RMF_Delta_selfcon2} in their ge\-neralized form become,
\begin{align}
 \label{eq:RMF_gen_selfcon1}
 \frac{\partial U}{\partial \sigma}&=\Gamma_{\sigma N} \sum_X \frac{m_X}{m_N} \rho_s^X(T,\mu^*_N,m_X^*),\\
 \label{eq:RMF_gen_selfcon2}
 \frac{\partial O}{\partial \omega}&=\Gamma_{\omega N} \sum_X \rho_B^X(T,\mu^*_N,m^*_X),
\end{align}
and the pressure reads
\begin{align}
 P=&\sum_X \left(P_0^X(T,\mu^*_N,m^*_X) + \Sigma^{0(r)}_X \rho_B^X  -\Sigma^{s(r)}_X\rho_s^X \right) \notag \\
 &-U(\sigma)+O(\omega).
\end{align}
The sum runs over all baryons that we include as interacting particles. These baryons have to be omitted in the HRG contribution
in case of the IHRG. We will only discuss the cases of interacting nucleons as well as interacting nucleons and $\Delta$'s in the
following. Another reasonable choice are all baryons in the spin-$1/2$ octet. However, the results are similar to the case of
interacting nucleons and $\Delta$'s since the masses of the $\Delta$, the $\Sigma$, $\Lambda$ and $\Xi$ are all
in the vicinity of $m \approx 1200 \ \text{MeV}$. The $\Delta$'s are 16-times degenerated, the $\Sigma$'s, $\Lambda$'s and
$\Xi$'s  have in total a degeneracy of 12, thus both cases are fairly similar, but the $\Delta$'s are more important in low-energy
heavy-ion collisions due to the reactions $\pi + N \leftrightarrow \Delta$.

We will now fix the parameters of the IHRG. The right-hand side of the selfconsistent equation \eqref{eq:RMF_gen_selfcon2} is
proportional to the net-baryon densities of the interacting baryons, which have to vanish for $\mu=0$. Since symmetries demand
that $O(\omega)$ is an even function, the left-hand side of the equation vanishes for $\omega=0$. This fixes $\omega=0$ for $\mu=0$ and
the repulsive interaction contributes only at finite chemical potential. As noted before this is different from repulsive
interactions introduced through excluded volume effects that contribute even at vanishing chemical potential
\cite{Rischke:1991ke,Yen:1997rv,Kapusta:1982qd,Hagedorn:1980kb}. We can therefore fix the scalar interaction solely with the lQCD
equation of state at $\mu_B=0$ and then tune the repulsive interaction to reproduce the nuclear equation of state at $T=0$ and
$\mu_B\neq 0$.

\begin{table}[t]
 \renewcommand{\arraystretch}{1.1}
	\begin{center}
		\begin{tabular}{|c|c|c|c|c|c|}
			\hline
	&	Int. baryons	& $\Gamma_{\sigma N}$ & $m_{\sigma} \ [\text{MeV}]$ & $B \ [1/\text{fm}]$  & $C$ \\ \hline
	NLDD1	&	$N$ & 28.64 & 550 & -29.67 & 3837\\ \hline
     NLDD2	&	$N+\Delta$ & 20.79 & 550 & -58.29 & 9690\\ \hline
		\end{tabular}
		\caption{Parameters for the scalar interaction in the IHRG at vanishing chemical potential $\mu_B=0$.}
		\label{tab:RMF_scalar}
	\end{center}
\end{table}

We use the following strategy to define the scalar interaction. We subtract the non-interacting HRG from the lQCD equation of
state and define in this way the contribution from the attractive mesonic interactions. We keep the scalar coupling as a constant
$\Gamma_{\sigma N}$, and obtain from Eq. \eqref{eq:coupling_gen_RMF} a constant ratio for the effective masses
$m^*_X/m^*_N=m_X/m_N$. The entropy density of the interacting model for a given temperature $T$ is then a function of only the
effective nucleon mass $m^*_N$,
\begin{equation}
 s_{\text{Int}}=s_0^N(T,m_N^*)+s_0^{\Delta}\left(T,\frac{m_{\Delta}}{m_N} m_N^*\right).
\end{equation}
We demand that the interacting entropy density $s_{\text{Int}}$ is equal to the missing entropy density (in the HRG) to reproduce
the lQCD result. This determines the effective mass $m^*_N(T,\mu_B=0)$. With $m^*_N(T)$ fixed we can easily calculate the scalar
densities and use the selfconsistent equation \eqref{eq:RMF_gen_selfcon1} to determine $\partial U/\partial \sigma(T)$ as a
function of temperature. The value of the $\sigma$-field as a function of temperature follows from the effective mass relation
$\sigma=(m_N-m^*_N)/\Gamma_{\sigma N}$. We can, furthermore,  fit $\partial U/\partial \sigma$ as a function of $\sigma$ and by
integration define the $\sigma$-selfinteraction. The polynomial ansatz for the selfinteraction (Eq. \eqref{eq:U_sigma}) is able to 
reproduce the interaction for both nucleons as well as nucleons and $\Delta$'s. The value of the scalar coupling $\Gamma_{\sigma 
N}$ is arbitrary, since $\sigma$ has no physical meaning, only $\Gamma_{\sigma N} \sigma=m_N-m^*_N$. If one rewrites the 
selfconsistent equations \eqref{eq:selfcon1} and \eqref{eq:RMF_gen_selfcon1} in terms of $m^*_N$ instead of $\sigma$, one finds 
for the polynomial ansatz of $U(\sigma)$ that the equation is determined by the ratios $m_{\sigma}/\Gamma_{\sigma N}$, 
$B/\Gamma_{\sigma N}^3$ and $C/\Gamma_{\sigma N}^4$. We fix $\Gamma_{\sigma N}$ by setting the $\sigma$-mass to its physical value 
$m_{\sigma} \approx 550 \ \text{MeV}$. The parameters in Tab. \ref{tab:RMF_scalar} give a good representation of the scalar 
selfinteraction for the temperature range between $T \approx 130 \ \text{MeV}$ and $T \approx 160 \ \text{MeV}$ and define the 
attractive interaction of the IHRG at $\mu_B=0$. We denote the set for interacting nucleons by 'NLDD1' and for nucleons and 
$\Delta$'s by 'NLDD2'. When comparing these parameters with those for other relativistic mean-field models one notices the large 
quartic coefficient $C$. Whereas the scalar selfinteraction in conventional mean-field models is just a small correction, it gives 
the dominant contribution in our approach. Another difference to conventional mean-field models is the much larger scalar density 
$\rho_s$ probed by the approach, since $\rho_s$ increases with $\rho_s \sim T^3$ for low $\mu_B$. This may lead to an unphysical 
phase transition if $\partial U/\partial \sigma$ is not strictly increasing monotonically, i.e. if the cubic or quartic 
coefficients $B$ or $C$ are negative. Note that the lQCD equation of state has no real phase transition and just a crossover, such 
that any model employed to describe lQCD cannot show a critical behavior for $\mu_B=0$. Both parametrizations, NLDD1 and NLDD2, 
have negative cubic interactions, but the dominant quartic interaction is positive such that the model is regular. In the same way 
- as a non-monotonic behavior in $\partial U/\partial \sigma$ can introduce a phase transition - also a non-monotonic behavior in 
$\partial O/\partial \omega$ potentially introduces one.

\begin{figure}[t]
	\centering
	\includegraphics[width=0.90\linewidth]{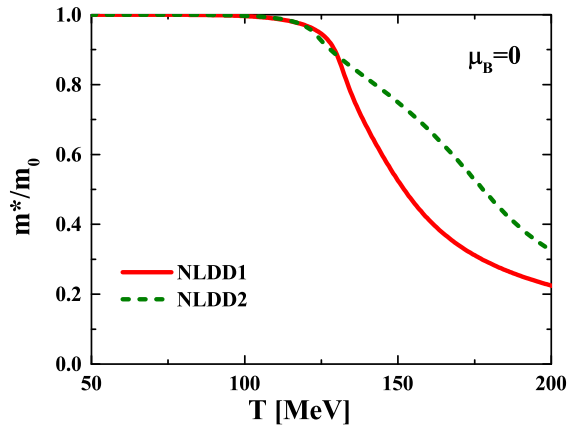}
	\caption{The effective baryon masses (scaled by their vacuum value) as a function of temperature $T$ for vanishing 
chemical potential. The full red line is the result with only interacting nucleons (NLDD1) and the dash-dotted green line for the 
case of interacting nucleons and $\Delta$-resonances (NLDD2).}
	\label{pic:VDD_meff_lat}
\end{figure}

We show in Fig. \ref{pic:VDD_meff_lat} the ratio of the effective masses (for the nucleon and $\Delta$) to the vacuum masses as a
function of the temperature for vanishing chemical potential. The additional interactions do not show up for temperatures below $T
\approx 100 \ \text{MeV}$, such that the effective masses stay at their vacuum values before decreasing with increasing
temperature. We note that a smaller effective mass results in a larger $\sigma$-field and more interaction strength compared to
the non-interacting case. The effective mass for the parameter set NLDD1 decreases more rapidly than for NLDD2, because the whole
additional interaction strength has to come from the nucleons alone while for NLDD2 also the $\Delta$-contribution is included.

\begin{figure}[t]
	\centering
	\includegraphics[width=0.90\linewidth]{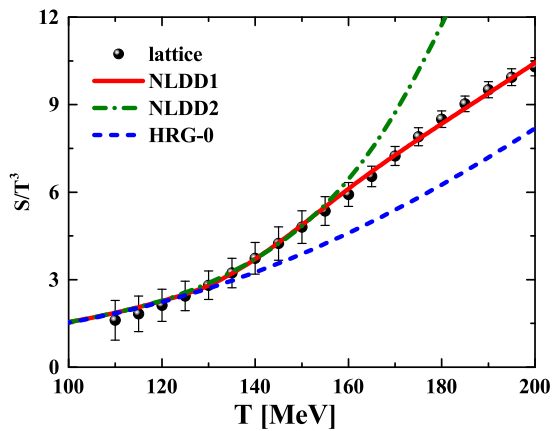}
	\caption{The scaled entropy density $s/T^3$ as a function of the temperature $T$ for vanishing chemical potential. The
		full red line is the entropy for the IHRG with interacting nucleons (NLDD1) and the dash-dotted green line for the 
IHRG with interacting nucleons and $\Delta$-resonances (NLDD2). The dashed blue line is the entropy density without interactions. 
The lQCD results are taken from the Wuppertal-Budapest collaboration \cite{Borsanyi:2013bia}.}
	\label{pic:VDD_S_comp}
\end{figure}

We show the corresponding scaled entropy densities $s/T^3$ in Fig. \ref{pic:VDD_S_comp} and compare them to the non-interacting 
HRG with the same degrees of freedom, HRG-0, and the lQCD entropy density from Ref. \cite{Borsanyi:2013bia} used to determine the 
attractive interaction. At small temperatures the interacting models are similar to the non-interacting HRG since the interactions 
give no contribution. The additional interaction becomes visible for $T \approx 125 \ \text{MeV}$. Up to temperatures of $T 
\approx 155 \ \text{MeV}$ both interacting models (by design) give the same result and describe the lQCD data within the error 
bars, however, differ at higher temperatures where one expects partonic matter anyhow. The IHRG with $\Delta$'s increases too 
fast for higher temperatures and exceeds the lQCD entropy while the model with only nucleons reproduces the entropy density even 
up to $T=200 \ \text{MeV}$. This is surprising since we fixed the interaction only for smaller temperatures. Nevertheless, the 
IHRG can not (and should not) describe the dynamics for temperatures beyond $T \approx 160 \ \text{MeV}$, since it does not use the 
proper degrees of freedom; however, both models work well in the region where we expect a hadronic medium.

\begin{figure}[t]
	\centering
	\includegraphics[width=0.90\linewidth]{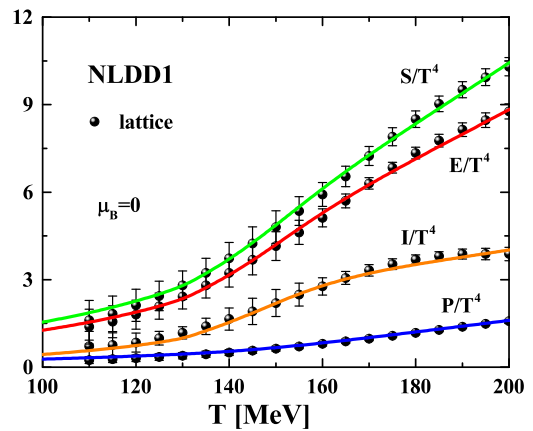}
	\caption{The equation of state for the IHRG for the parameter set NLDD1 as a function of the temperature $T$. The additional 
interaction is carried only by the nucleons. The green line shows the entropy density, the red line the energy density, the blue 
line the pressure and the orange line the interaction measure. All quantities are scaled by powers of the temperature. The results 
are within the error bars of the lQCD data in the whole temperature range displayed. The lQCD results are taken from the 
Wuppertal-Budapest collaboration \cite{Borsanyi:2013bia}.}
	\label{pic:I-HRG_EoS_N}
\end{figure}

\begin{figure}[t]
	\centering
	\includegraphics[width=0.90\linewidth]{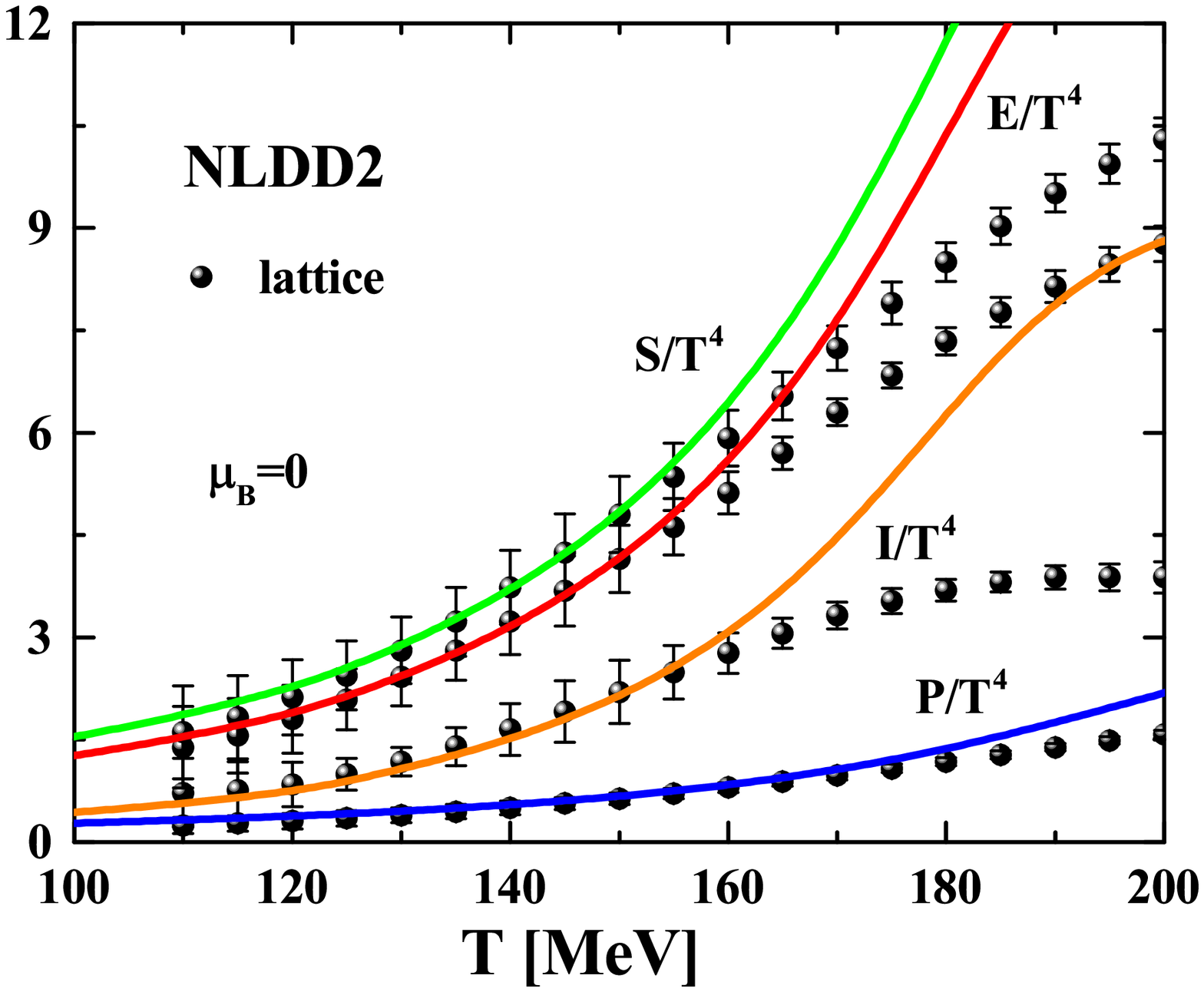}
	\caption{The equation of state for the IHRG for the parameter set NLDD2 as a function of the
		temperature $T$. The additional interaction is carried by the nucleons and the $\Delta$-resonances. The green line
shows the entropy density, the red line the energy density, the blue line the pressure and the orange line the interaction
measure. All quantities are scaled by powers of the temperature. The results are within the error bars of the lQCD data
for $T<160 \ \text{MeV}$. The lQCD results are taken from the Wuppertal-Budapest collaboration
\cite{Borsanyi:2013bia}.}
	\label{pic:I-HRG_EoS_ND}
\end{figure}

We now compare the equation of state from the two parametrizations with the lattice data from Ref. \cite{Borsanyi:2013bia} in
Fig. \ref{pic:I-HRG_EoS_N} (for NLDD1) and Fig. \ref{pic:I-HRG_EoS_ND} (for NLDD2) scaled by powers of the temperature. We find an
excellent agreement between lQCD and the model NLDD1, which describes the whole equation of state within the error bars of the data,
even at  temperatures $T > T_c$ where lQCD becomes more reliable and the error bars shrink. The thermodynamic consistency of the
approach ensures that we get the correct behavior in the pressure and the energy density once the entropy density is fixed. The
first differences will appear only for $T>200 \ \text{MeV}$. The interaction measure $I/T^4$ has its maximum around this
temperature and will then start to decrease in lQCD, but increases further in the hadronic model. The IHRG equation of state then 
exceeds the lQCD data. As mentioned earlier the parameter sets NLDD1 and NLDD2 give the same results for temperatures below $T=155 \ 
\text{MeV}$ and the model reproduces the lQCD data within the error bars for all temperatures below $T \approx 160 \ \text{MeV}$, 
where we expect a dominantly hadronic system (cf. Fig. \ref{pic:I-HRG_EoS_ND}). At larger temperatures the results from NLDD2 differ 
substantially from the lQCD results and rise too fast for $T>T_c$ as already seen in Fig. \ref{pic:VDD_S_comp}. One might improve 
the description at larger temperatures by using a different parametrization for the scalar selfinteraction and employ a larger 
temperature interval, but this is of no relevance for the present study where we have fixed the interaction only up to $T=160 \ 
\text{MeV}$. The excellent results from the parameter set NLDD1 at larger temperatures come out as a surprise.

\begin{figure}[t]
	\centering
	\includegraphics[width=0.90\linewidth]{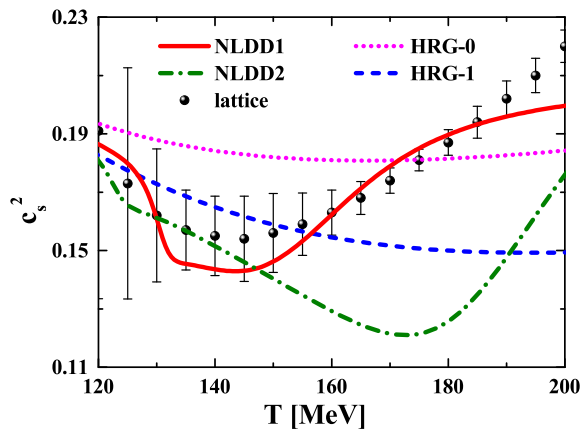}
	\caption{The speed of sound squared $c_s^2$ for the IHRG as a function of the temperature for vanishing chemical
		potential. The full red line is the result with interacting nucleons, the dash-dotted green line with interacting
nucleons and $\Delta$-resonances. The dotted magenta line line shows $c_s^2$ for the non-interacting HRG with all the hadrons in
the IHRG (HRG-0) and the dashed blue line for all hadrons listed by the particle data group \cite{Agashe:2014kda} with a mass
below $2.6 \ \text{GeV}$ (HRG-1). The lQCD results are taken from the Wuppertal-Budapest collaboration \cite{Borsanyi:2013bia}.}
	\label{pic:I-HRG_cs2}
\end{figure}

We have seen for the case of the conventional HRG that a reasonable reproduction of the equation of state does not imply a correct
behavior in the speed of sound, cf. Fig. \ref{pic:HRG_cs2_vs_T}. We now compare the speed of sound squared for the IHRG as a
function of the temperature in Fig. \ref{pic:I-HRG_cs2} and show again the corresponding results from the non-interacting HRG
(HRG-0) and also the HRG with hadrons up to a mass of $2.6 \ \text{GeV}$ (HRG-1). We find that only the parameter set NLDD1
describes the data properly. It reproduces the minimum at $T \approx 150 \ \text{MeV}$ and is within the error bars up to $T=180 
\ \text{MeV}$. Nevertheless, it benefits from the huge error bars at low temperatures. The parameter set NLDD2 can only describe the
data up to $T \approx 150 \ \text{MeV}$; it has also a minimum in $c_s^2$, but at a too high temperature, which is also too deep. 
On the other hand HRG-0 is completely off the data and has also the wrong $T$-dependence. The other version, HRG-1, with much more
particles gives a better description but only up to $T \approx 155 \ \text{MeV}$. From there on it has also the wrong slope. All 
models except for NLDD1 fail to describe the rise in the speed of sound at $T \approx 150 \ \text{MeV}$. For this it is necessary 
that the models reproduce the equation of state up to the inflection points of the scaled equation of state. Thus one will always 
find a decreasing speed of sound in $T$ if the scaled pressure $P/T^4$ has an increasing slope, which is a problem in most 
hadronic models.

\begin{figure}[t]
	\centering
	\includegraphics[width=0.90\linewidth]{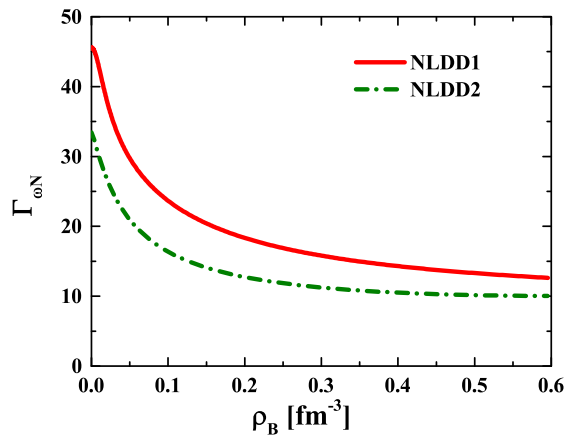}
	\caption{The density-dependent vector coupling $\Gamma_{\omega N}$ as a function of the net-baryon density $\rho_B$. The
full red line follows from the parameter set NLDD1, the dash-dotted green line from the set NLDD2. The energy density in Eq. 
\eqref{eq:VDD_gw_rho} is from our benchmark equation of state from Ref. \cite{Weber:1992qc}.}
	\label{pic:VDD_g_omega}
\end{figure}

With the scalar interaction defined at $\mu_B=0$ we can now discuss the repulsive interaction in addition. We fix it in the
same way as the scalar interaction using the nuclear equation of state at $T=0$ as input. In this limit the HRG contribution of
the IHRG vanishes and it reduces to a normal (density-dependent) relativistic mean-field model. The scalar interaction defines
already the effective masses as a function of the density at $T=0$ and therefore the selfinteraction $U(\sigma)$, the scalar
selfenergy $\Sigma^s$ and the non-interacting part $E_0$ of the energy density of the relativistic mean-field model
\eqref{eq:RMF_E}. The remaining contributions to the energy density depend on the repulsive interaction that we determine as
follows: We omit the selfinteractions in the $\omega$-field and keep only the mass term as in Eq. \eqref{eq:O_omega}, however, we
describe the repulsive interaction with a density-dependent vector coupling, which depends on the net-baryon density,
$\Gamma_{\omega N}(\rho_B)$. Note that it is important for the consistency of the model that $\rho_B$ contains only the interacting
baryons and not the whole baryon density of the IHRG; however, this is naturally the case for the nuclear equation of state. The
coupling can not depend on the scalar density, since this would lead to a scalar rearrangement selfenergy that alters the
effective mass and therefore the equation of state at finite temperatures and vanishing chemical potentials. The selfconsistent
equations for the $\omega$-field \eqref{eq:selfcon2}, \eqref{eq:RMF_Delta_selfcon2} and \eqref{eq:RMF_gen_selfcon2} simplify to
$m_{\omega}^2 \omega=\Gamma_{\omega N}(\rho_B) \rho_B$. We use this form of the $\omega$-selfinteraction to rewrite the equation
for the energy density \eqref{eq:RMF_E} as,
\begin{align}
\label{eq:VDD_E}
 E(T,\mu_B)-E_0(T,\mu^*,m^*)-U(\sigma)&=-O(\omega)+\Sigma^{0(0)} \rho_B \notag \\
 &=\frac{1}{2 m_{\omega}^2} \Gamma_{\omega N}^2(\rho_B) \rho_B^2.
\end{align}
The right-hand side of the equation depends on the repulsive interaction but the left-hand side is determined by the scalar
interaction and the equation of state that one wishes to reproduce. Since both are already defined, we can use Eq. 
\eqref{eq:VDD_E} to determine the vector coupling $\Gamma_{\omega N}$.
\begin{table}[t]
	\renewcommand{\arraystretch}{1.3}
	\begin{center}
		\begin{tabular}{|c|c|c|}
			\hline
			&	NLDD1	& NLDD2 \\
			\hline
			Int. baryons & N & N+$\Delta$ \\
			\hline
			$\tilde{A}$ & 45.59 & 33.44 \\
			\hline
			$\tilde{B}$ & 3045 & 50231 \\
			\hline
			$\tilde{C}$  & $4.90 \cdot 10^7$ & $1.99 \cdot 10^7$\\
			\hline
			$\tilde{D}$  & $1.40 \cdot 10^{10}$ & $-2.75\cdot 10^9$ \\
			\hline
			$\tilde{E}$  & $6.21 \cdot 10^7$ & $1.18 \cdot 10^8$ \\
			\hline
			$\tilde{F}$  & $7.63 \cdot 10^{10}$ & $-1.49\cdot 10^{10}$ \\
			\hline
		\end{tabular}
		\caption{Parameters for the vector interaction in the IHRG at vanishing temperature. All parameters are in units
			of GeV.}
		\label{tab:VDD_coeff}
	\end{center}
\end{table}
Note that we describe the equation of state at $T=0$ as a canonical system, where the density is a natural variable (instead of
the chemical potential) and the chemical potential follows as a derivative of the thermodynamic potential with respect to density.
Therefore, Eq. \eqref{eq:VDD_E} defines the coupling directly as a function of the density $\rho_B$,
\begin{equation}
\label{eq:VDD_gw_rho}
 \Gamma_{\omega N}(\rho_B)=\sqrt{2} \ \frac{m_{\omega}}{\rho_B}
\sqrt{E(\rho_B)-E_0(\rho_B,m^*)-U(\sigma)},
\end{equation}
where the repulsive interaction is actually determined by the ratio $\Gamma_{\omega N}/m_{\omega}$. This is similar to the scalar
interaction where we fixed $m_{\sigma}$ to its physical value. We do the same here and fix $m_{\omega}=783 \ \text{MeV}$. With Eq.
\eqref{eq:VDD_gw_rho} one can employ any possible nuclear equation of state. Even if it is possible to directly use the function
in
the numerical calculations, it is convenient to have a parametrized form. We will use the ansatz
\begin{equation}
\label{eq:VDD_gw_fit}
\Gamma_{\omega N}(\rho_B)=\tilde{A} \cdot \frac{1+ \tilde{B} |\rho_B|+ \tilde{C} |\rho_B|^2+ \tilde{D} |\rho_B|^3}{1+\tilde{B}
|\rho_B| + \tilde{E} |\rho_B|^2+\tilde{F} |\rho_B|^3}.
\end{equation}
This function is similar to the ansatz employed in Refs. \cite{Typel:1999yq,Hofmann:2000vz} to fit the density-dependence
of Dirac-Brueckner calculations, but with two important differences: The model in Refs. \cite{Typel:1999yq,Hofmann:2000vz} was 
only applied to nuclear matter and finite nuclei, i.e. to $T=0$, but we want to employ our model also at finite temperatures and 
vanishing chemical potential. It is therefore mandatory that the coupling is an even function of the density, which is guaranteed 
by the absolute values of the density that lead to $\Gamma_{\omega N}(\rho_B)=\Gamma_{\omega N}(-\rho_B)$, but the function has to be 
continuously differentiable at $\rho_B=0$. We ensure this by taking the same linear coefficient in the numerator and denominator 
of Eq. \eqref{eq:VDD_gw_fit}. To compensate for the lost coefficient in the linear term we extend the polynomials in the function 
to  third order. The coefficients for the fit are summarized in Tab. \ref{tab:VDD_coeff} and we show the couplings as a function 
of $\rho_B$ in Fig. \ref{pic:VDD_g_omega}. It might not be possible to see by eye, but both functions fulfill the condition 
$\Gamma_{\omega N}'(\rho_B=0)=0$. The differences in the couplings are not due to the appearance of $\Delta$'s but follow from the 
different scalar interactions. Since the parameter set NLDD1 has a stronger scalar, i.e. attractive interaction, it needs a 
stronger repulsive interaction to balance it.

\begin{figure}[t]
	\includegraphics[width=0.90\linewidth]{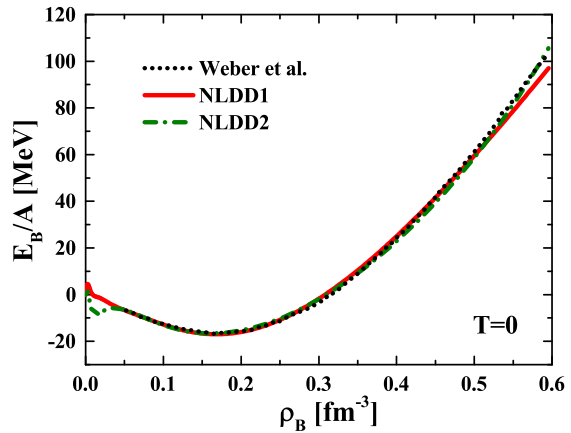}
	\caption{The binding energy per nucleon $E_B/A$ as a function of the nucleon density $\rho_B$ for $T=0$. The dotted black
		line is taken from Weber {\it et al.} \cite{Weber:1992qc} and is the benchmark for our nuclear equation of state. The 
full red line is the result for the parameter set NLDD1 and the dash-dotted green line for the set NLDD2.}
	\label{pic:VDD_nuc_EoS}
\end{figure}

We can now calculate the nuclear equation of state of the IHRG to check the quality of the fit \eqref{eq:VDD_gw_fit}. The binding
energy per nucleon $E_B/A$ is shown in Fig. \ref{pic:VDD_nuc_EoS} in comparison to the result from Weber {\it et al.}
\cite{Weber:1992qc}, which we use as a benchmark for the nuclear equation of state. All three lines are approximately on top of
each other up to $\rho_B = 3 \rho_0$; thus the vector coupling is successful in reproducing the binding energy. Both
parameter sets (NLDD1 and NLDD2) give nuclear binding energies in the range of $E_B=-16$ to $-17 \ \text{MeV}$ and a saturation
density of $\rho_0=0.168 \ \text{fm}^{-3}$ for NLDD1 and $\rho_0=0.161 \ \text{fm}^{-3}$ for NLDD2. Note that the nuclear equation of
state for NLDD2 gets no contribution from the interacting $\Delta$'s. Even if their mass gets reduced due to the scalar interaction,
they are still too heavy to give a thermodynamic contribution. The same holds also if one includes even more interacting hadrons,
since they would additionally weaken the scalar interaction. The nuclear equation of state of the IHRG is therefore predominantly
determined by the nucleons. Fig. \ref{pic:VDD_nuc_EoS} shows some odd behaviors at very small densities: NLDD1 has a very small peak
and NLDD2 exhibits a second minimum at $\rho_B=0.012 \ \text{fm}^{-3}$ that is higher than the global minimum at saturation density
implying that it describes just a metastable state. These deviations can appear also in conventional relativistic mean-field
models, but at even smaller densities. They occur due to the singular behavior of the binding energy per nucleon
$E_B/A=E/\rho_B-m_N$, if the energy density differs from the form $E \approx m_N \cdot \rho_B$ at small densities. In case of the
IHRG they result from the condition $\Gamma_{\omega N}'(\rho_B=0)=0$ that constrains the functional form of
$\Gamma_{\omega N}(\rho_B)$. However, the effects are small and negligible if one considers systems with energy densities above 20
MeV/fm$^3$.	

\begin{figure}[t]
	\centering
	\includegraphics[width=0.90\linewidth]{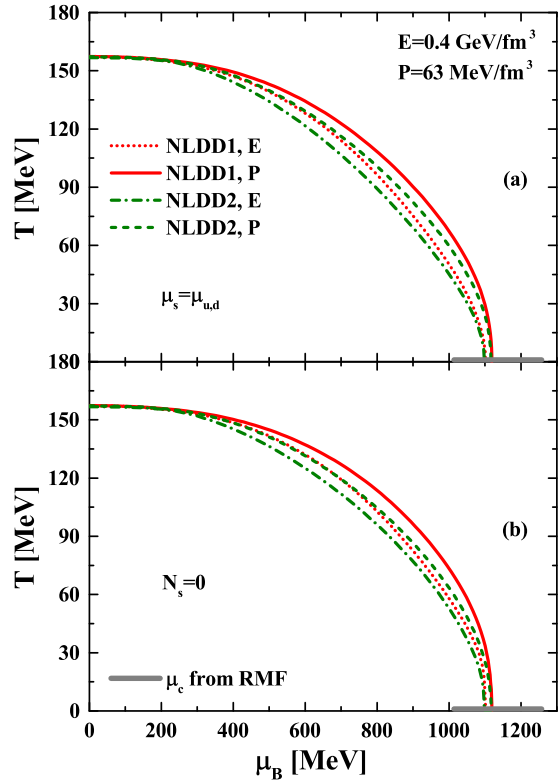}
	\caption{Lines of constant pressure and energy density from the IHRG in the $(T, \mu_B)$-plane without $(a)$ and with 
strangeness neutrality $(b)$. A constant energy density of $E=0.4 \ \text{GeV}/\text{fm}^3$ is shown by the dotted red
line for the parameter set NLDD1 and by the dash-dotted green line for the set NLDD2. A constant pressure of $P=63 \ 
\text{MeV}/\text{fm}^3$ is shown by the full red line for the set NLDD1 and by the dashed green line for the set NLDD2. The grey band 
(at $T \approx 0$)  highlights the ambiguity of the chemical potential at zero temperature for relativistic mean-field 		
theories with different repulsive interactions.}
	\label{pic:PD_strange}
\end{figure}

So far we have shown that the IHRG gives reasonable results for $\mu_B=0$ along the $T$-axis as well as for low $T$ and finite
nuclear density. However, the IHRG is defined in the whole $(T, \mu_B)$ plane and we may explore the phase diagram e.g. for
constant thermodynamical potential, which for a grand-canonical ensemble is the (negative) pressure $P$. Alternatively, for $T
\rightarrow$ 0 the thermodynamic potential is the energy density $E$. We show in Fig. \ref{pic:PD_strange} the lines of constant
pressure $P=63 \ \text{MeV}/\text{fm}^3$ and of constant energy density $E=0.4 \ \text{GeV}/\text{fm}^3$, which are roughly the
values of the lQCD equation of state for $T_c \approx 155 \ \text{MeV}$ and vanishing chemical potential. Accordingly, the lines 
may be interpreted as a QCD phase boundary. Both conditions give similar results but the lines of constant pressure reach further 
out into the $(T, \mu_B)$ plane. When addressing the phase boundary in the context of heavy-ion collisions, an important 
constraint is strangeness neutrality. In Fig. \ref{pic:PD_strange} (a) the lines of constant pressure or energy density were 
obtained without any constraint on the strange sector, such that we have a finite strangeness $N_S >0$, while we show in Fig. 
\ref{pic:PD_strange} (b) the lines for a strange neutral medium. However, neither strangeness neutrality nor the two different 
parametrizations of the IHRG, NLDD1 and NLDD2, have a strong impact on the phase boundaries, which agree close to the axis of the 
$(T, \mu_B)$ plane and cover almost the same area in the phase diagram.

We note in passing that a large uncertainty in the phase boundary stems from the transition at low temperatures ($T \approx 0$) 
and large baryon chemical potentials. As mentioned earlier the nuclear equation of state is known only as a function of the
density $\rho_B$, but not of the chemical potential $\mu_B$. As mentioned above the strength of the repulsive interaction between
the nucleons has a large influence on $\mu_B$ since the thermodynamics are defined by the effective baryon chemical potential
$\mu^*$ that is shifted by the vector selfenergy \eqref{eq:mueff}. Thus models with different interactions can reproduce the same
nuclear equation of state as a function of density $\rho_B$, but for different chemical potentials. The grey band at $T \approx 0$
in Fig. \ref{pic:PD_strange} indicates the uncertainty for small temperatures. The lowest chemical potential is the estimate for 
a gas of non-interacting nucleons, the largest for the parameter set NL3 from Ref. \cite{Lang:1991qa}, which has a very strong 
repulsive interaction but is consistent with the ground-state properties of nuclear matter.

\begin{figure}[t]
	\centering
	\includegraphics[width=0.90\linewidth]{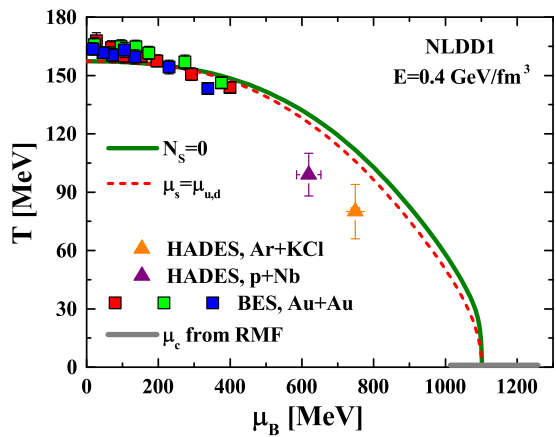}
	\caption{Lines of constant energy density from the IHRG in the $(T, \mu_B)$-plane for the parameter set NLDD1 with (solid 
green) and without (dashed red) strangeness neutrality. The grey band (at $T \approx 0$) highlights the ambiguity of the chemical 
potential at zero temperature for relativistic mean-field theories with different repulsive interactions. The squares are 
freeze-out points taken from Ref. \cite{Adamczyk:2017iwn} for $0$-$5 \%$ (red), $30$-$40 \%$ (green) and $60$-$80 \%$ (blue) 
centrality. The triangles are freeze-out points taken from Ref. \cite{Agakishiev:2015bwu}.}
	\label{pic:PD+freezeout}
\end{figure}

To quantize the effects of strangeness neutrality we show in Fig. \ref{pic:PD+freezeout} the lines of constant energy density 
calculated for NLDD1 with (solid green line) and without (dashed red line) strangeness neutrality. The lines agree close to the 
axis of the $(T,\mu_B)$-plane because one has the same amount of strange and antistrange particles at $\mu_B=0$ and no strange 
particles at all at $T=0$. These regions of the phase diagram are always strange neutral and therefore have to agree independent 
of the strange sector while for $T \neq 0 \neq \mu_B$ we find that the phase boundaries for a strange neutral medium are shifted 
further into the $(T, \mu_B)$-plane because
strangeness neutrality reduces the amount of particles in the strange sector compared to an unconstrained system. This lowers the 
energy density and the pressure and one needs larger temperatures and baryon chemical potentials to reach the same energy 
densities and pressures as in an unconstrained medium.

We also compare our predictions for the phase boundary to recent results for freeze-out points from the beam-energy-scan (BES) 
program at RHIC \cite{Adamczyk:2017iwn} and from the HADES collaboration \cite{Agakishiev:2015bwu}. The freeze-out data from BES 
are all at small chemical potentials and located around the predicted phase boundary, the freeze-out data from HADES are at larger 
baryon chemical potentials and lower than the IHRG boundary. This is expected since in heavy-ion collisions below about 2 AGeV no 
'droplets' of QGP can be created and the system entirely stays in the hadronic phase \cite{Linnyk:2015rco}.

\section{Summary}
\label{sec:summary}

In this work we have discussed QCD thermodynamics in the regime of hadronic degrees of freedom. At finite temperature $T$ and 
vanishing chemical potential $\mu_B$ the interactions are dominated by resonant scatterings and the thermodynamics  can  well be 
described by a HRG model where one includes the resonances as non-interacting particles in the partition sum. The approach can 
reproduce the equation of state, as provided by lQCD, but requires a large amount of interaction strength, i.e. hadronic 
resonances. The HRG describes only the attractive interactions between the hadrons while short-range repulsive interactions, that 
are important at large densities, have to be introduced by means of excluded-volume models. These interactions, however, are only 
thermodynamically motivated and not based on field theory. Hence they are in general not covariant and thus can not be used in 
transport approaches for non-equilibrium configurations.

A covariant formulation for repulsive (short-range) interactions is provided by relativistic mean-field theories (RMFTs). These 
approaches describe the interactions by meson exchange and characterize hadronic matter at low temperatures. Consequently, the 
model is a well-known and suitable approach for nuclear matter at finite density and successful in describing the ground-state 
properties of the nuclear equation of state. In particular, we have discussed a relativistic mean-field approach for symmetric 
nuclear matter with density-dependent couplings. This extension allows for a  parametrization of  realistic nucleon-nucleon 
interactions - e.g. from Dirac-Brueckner calculations - and introduces a non-trivial density dependence in the scalar and vector 
interactions.

By combining both hadronic models, the HRG and RMFT, we have constructed an interacting HRG, denoted by IHRG, that reproduces the 
lQCD equation of state at $\mu_B \approx 0$, $T>0$ and the nuclear equation of state at $T \approx 0$, $\mu_B>0$. The $s$-channel 
interactions are described - as in the HRG - by the inclusion of resonances as non-interacting particles; the $t$ channel, i.e. 
meson-exchange reactions, are treated by means of RMFT. The repulsive interactions in the IHRG are 
therefore fundamentally different from those incorporated in excluded-volume models, which are proportional to the total particle 
number or the pressure. In the IHRG the repulsive interaction is controlled by the net-baryon density and vanishes for $\mu_B=0$. 
The attractive interactions are realized by  different mechanisms and it is therefore not necessary to include a huge amount of 
hadronic resonances in the IHRG since one can focus on well known hadronic states and resonances; the missing interaction strength 
is provided by the $\sigma$-meson exchange.

We have presented two parametrizations of the IHRG: In the first, denoted by NLDD1, we have only considered the meson-exchange 
reactions for nucleons. In the second, denoted by NLDD2, we have included additionally the $\Delta$-resonances, which are important 
for the dynamics of heavy-ion collisions in the pion-nucleon channel. We have defined the interactions of the $\Delta$'s in such a 
way  that the ratio of the effective masses is given by the ratio of the vacuum masses but the effective chemical potentials for 
nucleons and $\Delta$'s are the same. Using this approximation one can easily extend the IHRG to many interacting baryons, 
however, the mesons are kept non-interacting.

Since the IHRG can describe the nuclear equation of state at $T=0$ and finite density,  it should give a better description of 
the hadronic medium at low temperatures and finite density than the standard HRG. This region of the phase diagram is of 
particular interest for heavy-ion collisions at low beam energies that investigate the phase boundary of QCD at large baryon 
chemical potentials and search for a possible critical point in the phase diagram. We have given an estimate for the QCD phase 
boundary between the QGP and hadronic configurations based on the assumption that the phase transition occurs at a constant energy 
density or pressure. We have found that both conditions lead to almost the same transition region for both sets, NLDD1 and NLDD2.
The predicted boundary is consistent with the freeze-out analysis from the BES that probed the hot QCD medium at moderate chemical 
potentials. We, furthermore, note that the phase boundary has a large uncertainty at low temperatures since the nuclear equation 
of state is only known as a function of the density (canonical ensemble) and not the chemical potential (grand-canonical 
ensemble).

\section*{Acknowledgments}

The authors acknowledge valuable discussions with E. L. Bratkovskaya and M. Gorenstein  during this study which was supported by 
the Bundesministerium f\"ur Bildung und Forschung (BMBF).

\end{document}